# Deep Reinforcement Learning for Active Flow Control around a Circular Cylinder Using Unsteady-mode Plasma Actuators

M. A. Elhawary[a]

a Independent Researcher, 12613 Giza, Egypt

## Abstract

Deep reinforcement learning (DRL) algorithms are rapidly making inroads into fluid mechanics, following the remarkable achievements of these techniques in a wide range of science and engineering applications. In this paper, a deep reinforcement learning (DRL) agent has been employed to train an artificial neural network (ANN) using computational fluid dynamics (CFD) data to perform active flow control (AFC) around a 2-D circular cylinder. Flow control strategies are investigated at a diameter-based Reynolds number $Re_D = 100$ using advantage actor-critic (A2C) algorithm by means of two symmetric plasma actuators located on the surface of the cylinder near the separation point. The DRL agent interacts with the computational fluid dynamics (CFD) environment through manipulating the non-dimensional burst frequency ($f^+$) of the two plasma actuators, and the time-averaged surface pressure is used as a feedback observation to the deep neural networks (DNNs). The results show that a regular actuation using a constant non-dimensional burst frequency gives a maximum drag reduction of 21.8 %, while the DRL agent is able to learn a control strategy that achieves a drag reduction of 22.6%. By analyzing the flow-field, it is shown that the drag reduction is accompanied with a strong flow reattachment and a significant reduction in the mean velocity magnitude and velocity fluctuations at the wake region. These outcomes prove the great capabilities of the deep reinforcement learning (DRL) paradigm in performing active flow control (AFC), and pave the way toward developing robust flow control strategies for real-life applications.

Corresponding author: Mohamed Elhawary maaelhwary@gmail.com

# 1. Introduction:

Bluff-body flow control has received increasing attention in the recent years due to the large amount of energy that is being consumed annually in overcoming aerodynamic forces encountered in many engineering applications such as ground, air, and sea vehicles [1]. In ground vehicles, for example, the aerodynamics drag consumes approximately 27% of the total fuel consumption [2], while a minor drag reduction could result in sizeable fuel-savings. In this context, It has been estimated that around 3.5-billion dollars could be saved from reducing only 10% of the total drag encountered by a commercial fleet of airplanes in the United States alone [1]. Thus, achieving effective and stable flow control strategies is of a prominent importance.

Flow past cylinder offers excellent opportunities to study flow control strategies because it exhibits several fluid dynamics phenomenon, such as large-scale flow separation and alternating vortex shedding. Once a flow control strategy proved effectiveness in controlling flow separation around a cylinder, it can be easily translated to general geometric configurations [3]. Several techniques have been investigated to study flow control including passive, active and reactive methods [1]. Passive control methods such as vortex generators [4], spoilers [5], and flaps [6], have been the centerpiece of flow control for a long time thanks to their ease of use. However, such methods have inherent drawbacks, especially their narrow operating range and their negative effects at off-design conditions [7]. To overcome these drawbacks, active flow control (AFC) methods have been introduced due to their advantages of being adaptive to the flow conditions as they can be switched off when flow control is not necessary. Active flow control techniques include: steady suction/blowing [8], acoustic excitation [9], synthetic jets [9], and plasma actuators [10]. Among all these techniques, plasma actuators have received particular attention due to their fast response, their high repetition rate, and their ability to produce zero-mass flux with no moving parts [3]. A drag reduction of approximately 25% is reported in [10] from controlling the flow separation around a circular cylinder using two symmetric plasma actuators at Reynolds number of 10000. In addition to flow control, plasma actuators demonstrate great capabilities in other control tasks such as noise-reduction [11], and transition-delay [12].

Despite the remarkable results from active flow control (AFC), achieving effective and stable flow control strategies is as a highly sophisticated process due to the nonlinearity of fluid dynamics and the high dimensionality of the control parameters [13]. For this purpose, machine learning (ML) algorithms (i.e., data-driven optimization and applied regression) have been utilized to discover flow control strategies. ML algorithms offer a completely different paradigm that is able to reach quicker solutions without a prior information about the problem physic [14]. Based on the information that is available for training the learning machine, ML algorithms can be classified into three main categories: supervised, semi-supervised, and unsupervised learning [14], [15]. Genetic algorithms (GAs) [16] are the first machine learning techniques that have been utilized for studying flow control problems. Recently, reinforcement learning (RL) has been attracting significant attention due to its intriguing capabilities in performing complex control tasks in several fields of science and engineering [17]. In a typical reinforcement learning (RL) algorithm, an agent interacts with an environment and receives a reward corresponding to an action, then the agent learns from experience to take actions that maximize the expected cumulative reward. A major breakthrough in RL research has been achieved after the integration between reinforcement learning (RL) and deep neural networks (DNNs), for what is called deep reinforcement learning (DRL). This integration reveals sever obstacles that hinder the vanilla RL algorithms, and helps achieving remarkable results in complex problems such as robotics control [17], and beating professional players in the game of Go [18].

The implementation of reinforcement learning algorithms in fluid mechanics applications include shape optimization [19], fish bio-locomotion [20], passive flow control [21], and active flow control [13], [22], [23]. The pioneering work by Jean Rabault *et al.* [13] presents the first application of deep reinforcement learning (DRL) algorithms for active flow control (AFC) based on computational fluid dynamics (CFD) data. In this study [13], the authors achieved a drag reduction around 8% from controlling flow separation around a 2-D circular cylinder at Reynolds number of 100 using two synthetic jets by means of proximal policy optimisation (PPO) method [24]. H. Tang *et al.* [23] revisited the same DRL-based AFC problem in Jean Rabault *et al.* [13] at a higher Reynolds number, up to 400, using four synthetic jets and the authors reported a drag reduction of approximately 5.7%, 21.6%, 32.7%, and 38.7% for Reynolds numbers 100, 200, 300, and 400, respectively. The current study revisit the same DRL-based AFC problem in [13], [23] using a relatively new active control technique, namely plasma actuators [25], by means of a different deep reinforcement learning (DRL) method called advantage actor-critic (A2C) algorithm [26].

In this study, active flow control around a 2-D circular cylinder is investigated by means of the advantage actor-critic (A2C) algorithm via manipulating the non-dimensional burst frequency ($f^+$) of two symmetric plasma actuators. Four non-dimensional burst frequencies $f^+$ = 0.5, 1.0, 1.5, and 2.0, corresponding to plasma burst frequencies $f_b$ = 0.08 Hz, 0.16 Hz, 0.24 Hz, and 0.32 Hz, are tested at a fixed actuation body force ($F$) and constant duty-cycle ($\tau$). Numerical simulations are carried-out to solve the 2-D unsteady incompressible Navier-Stokes equations, at a diameter-based Reynolds number $Re_D$ = 100, using incremental pressure correction scheme (IPCS). A detailed comparison between the uncontrolled case (without actuation) and the controlled case is presented, and the effectiveness of the control strategy is evaluated in terms of the improvement in the drag coefficient, the lift coefficient, and the mean velocity at the wake region. To the author best knowledge, this is the first study that discusses the use of the A2C algorithm in active flow control problems; thus, all codes are released as open-source to facilitate the further development.

## 2. Methodology:

### 2.1. Problem description:

The computational domain used in the current study is set similar to the well-known benchmark in [27]. All quantities are put in a non-dimensional form with respect to the diameter of the cylinder ($D$). As shown in Figure 1, the circular cylinder is immersed in a rectangular box with a non-dimensional length of $22D$ and a non-dimensional height of $4.1D$.

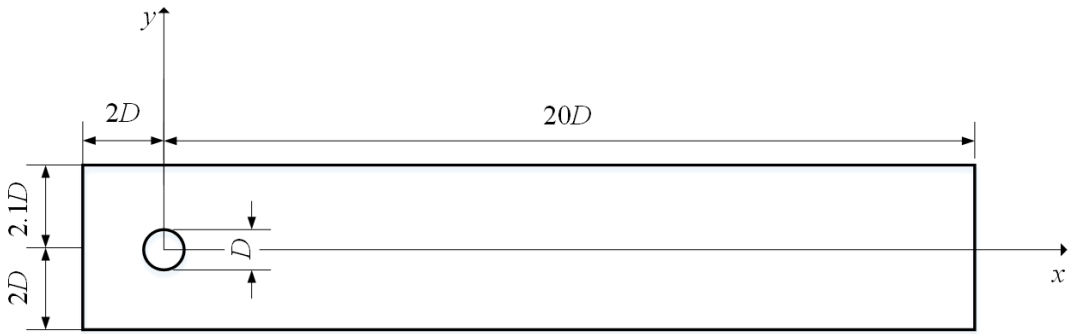

Figure 1: Geometrical description of the computational domain

A no-slip boundary condition is set at the top and the bottom walls as well as on the solid surface of the cylinder. The vertical component of the velocity at the inlet boundary is set as $v_{in} = 0$, and the axial component ($u_{in}$) is prescribed as a parabolic profile following the formula:

$$u_{in}(0, y) = 4u_m y(h - y)/h^2, \qquad (1)$$

in which $u_m, y,$ and $h$ are the mean velocity, the vertical distance from bottom wall, and the height of the inlet boundary, respectively. The outlet boundary condition is considered as zero pressure with the assumption of zero velocity gradient (i.e., fully developed velocity profile), as follows:

$$pn - \mu \nabla u.n = 0, \qquad (2)$$

where $p, u, \mu,$ and $n$ are the pressure, the velocity, the dynamic viscosity, and the unit vector normal to the boundary, respectively. Similarly to the benchmark [27], the cylinder origin is shifted slightly in the vertical direction by 0.05D to help in triggering the vortex shedding.

### 2.2. Configuration of plasma actuators:

The typical plasma-actuator, shown in Figure 2, consists of two electrodes: the lower electrode is covered by a thin film of a dielectric layer, while the upper electrode is exposed to the air [28]. When applying enough power from an AC power supply at a reasonable frequency, the air around the upper electrode is ionized, and plasma is formed [28]. As a result, the plasma actuator induces a body force on the surrounding air in the form of a wall-jet downstream the upper electrode. The plasma actuators that are used in the current study are positioned at an angle $\theta = \pm 115°$ and they spread over the surface of the cylinder by $\theta_{plasma} = 7°$, as shown in Figure 3. The optimal location of the plasma actuators is studied intensively in the literature [29]–[31], and generally, the plasma actuators are recommended to be positioned near the flow-separation point. In the current study, the separation point is estimated between $\theta = \pm 110°$ and $\theta = \pm 115°$ from the front stagnation point, which is in a good agreement with the results reported in [32].

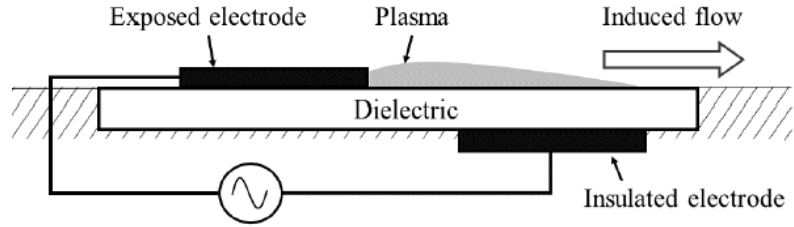

Figure 2: Schematic view of plasma actuator.

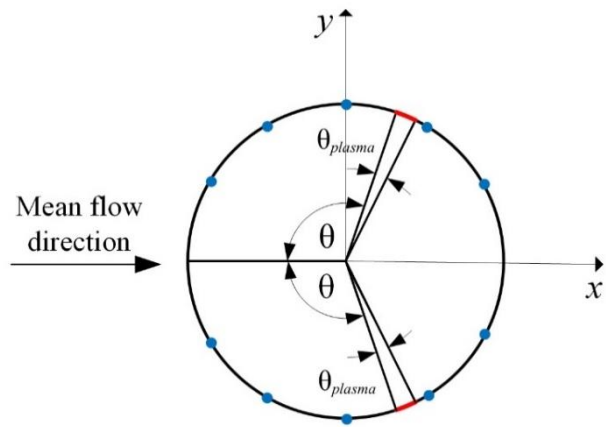

Figure 3: Description for the locations of the two plasma actuators and the ten pressure probes that are utilized for performing active flow control (AFC) around the circular cylinder. The plasma actuators (marked in red) are centered at $\theta = \pm 115°$, and they spread over the surface by $\boldsymbol{\theta_{plasma}} = 7°$. The pressure probes (marked in blue) are positioned at $\boldsymbol{\theta} = \pm 30°, 60°, 90°, 120°,$ and $150°$ from the first stagnation point.

Generally, plasma actuators have two modes of operation: steady mode and unsteady mode (also called burst mode). As shown in Figure 4, under steady mode the plasma actuator works at a continuous alternate body force, while under unsteady mode the actuator is controlled by a periodic on-off force. Previous studies discussed that actuation under unsteady mode can provide significantly greater aerodynamic performance than steady actuation [28], [33]. It has been also proved that the power requirement for the unsteady mode is less than the steady mode, while the gain could be the same or even better than the steady actuation as discussed in [28], [31], [34]. The actuation under unsteady mode is characterized by the duty-cycle $\tau = T_{on}/T_b$, in which $T_{on}$ and $T_b$ are the time of active modulation and the total time of burst wave, respectively. Another important parameter is the non-dimensional burst frequency $f^+ = f_b D/U$, in which $f_b, D,$ and $U$ are the burst frequency, the cylinder diameter, and the mean velocity magnitude. In the current study, the specifications of the two symmetric plasma actuators (i.e., the dimensions and the working parameters) are set similar to the plasma actuators that have been used in the experimental study of [3] in which the authors tested the plasma actuators in the range of $f^+ = 0.1 - 2.0$ and they measured a value of the body force between $F = 0.12 – 1.28$ mN.

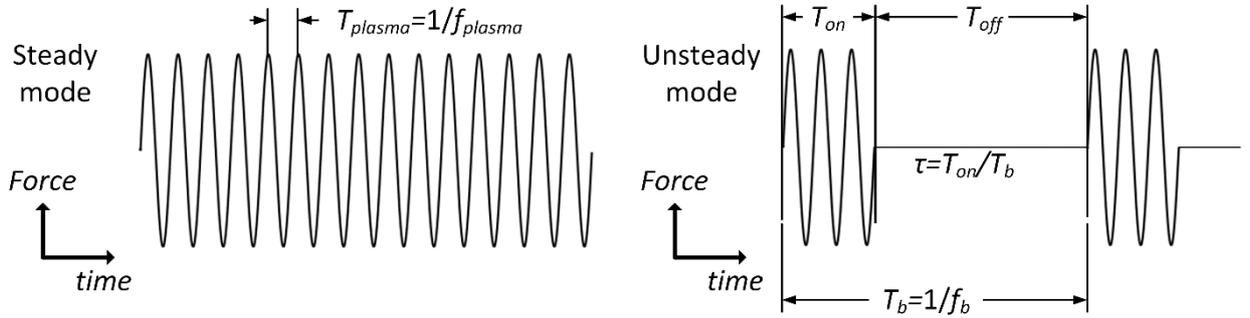

Figure 4: Waveforms of steady mode (left) and unsteady mode (right)

Several models have been developed to express the effect of the body force that is induced by the plasma actuators [35]–[37]. Suzen et al. [36] proposed a model derived from Maxwell'equations that expresses the body force in terms of the electric field and the charge density that are generated by the plasma actuator. In this approach, the body force is calculated by solving the partial differential equations of the Suzen-Huang's model [36], and then incorporate the value of the body force into the Navier-Stokes equations as a momentum source term. Although this method [36] is reported to be accurate, it is considered to be computationally expensive as it requires solving additional partial differential equations, which is not recommended in the current study due to the large amount of computations that are performed during the training of the deep neural networks (DNNs). An alternative simple and time-efficient approach is adopted by T. West et al. [34], in which the authors considered a value of the body force that is measured from an experimental study [38]; similarly, in the current study a value of the body force per unit volume $F = 0.7$ mN/m$^3$ from the experimental work of [3] is adapted. 3

Four non-dimensional burst frequencies, within the same range that is discussed in [3], namely $f^+= 0.5, 1.0, 1.5,$ and $2.0$ are tested at a constant duty-cycle $\tau = 85\%$. It is worth noting that the operating frequencies of the plasma actuators $(f_{plasma})$ are usually much higher than any fluid frequencies; thus, the flow is considered to receive a constant body force during each pulse (i.e., quasi-steady approximation) as discussed in [3], [39]. Figure 5 represents an example of the unsteady actuation body force that has been tested in the current study at a duty-cycle $\tau = 85\%$ and a non-dimensional burst frequency $f^+ = 0.5$, corresponding to a burst frequency $f_b = 0.08$ Hz. The figure also shows that a gradual rise and decay, for a period time equals to $0.2\ T_b$, is introduced at the beginning and end of each pulse to avoid the sudden change during the numerical solution, as suggested by [40].

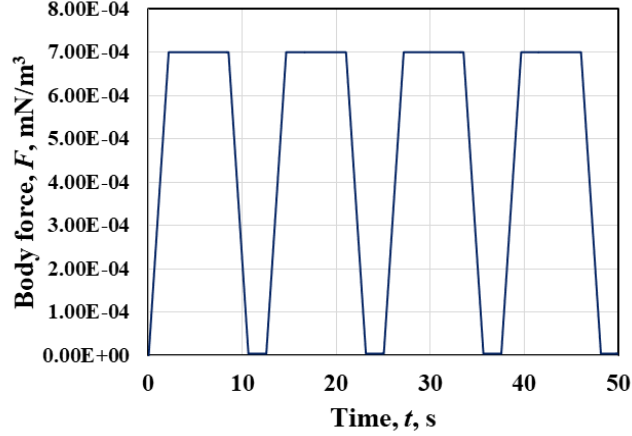

Figure 5: Graph of actuation body force presenting four unsteady actuation cycles at a duty-cycle $\tau = 85\%$ and a non-dimensional burst frequency $f^+ = 0.5$, corresponding to a burst frequency $f_b$ 0.08 Hz.

### 2.3. Numerical methods:

The flow field is described by the 2-D unsteady incompressible Navier-Stokes equations, and the actuation body-force that is generated from the interaction between air and the plasma actuators is incorporated as a momentum source term, as discussed in the previous section. The continuity and the momentum equations can be expressed as follows [41]:

$$\nabla . u = 0 \tag{3}$$

$$\rho \left(\frac{\partial u}{\partial t} + u . \nabla u\right) = -\nabla p + \nabla . \mu (\nabla u + \nabla u^T) + F, \tag{4}$$

where $u, p, t, \mu$, and $F$ are the velocity, the pressure, the time, the dynamic viscosity, and the body force per unit volume, respectively. The Reynolds number based on the diameter of the cylinder is defined as $Re_D = UD/\nu$, in which $D$ and $\nu$ are the cylinder diameter and the air kinematic viscosity, $U = 0.016 \, m/s$ is the mean velocity magnitude calculated as follows [23]:

$$U = \frac{1}{h} \int_{-2D}^{2.1D} u_{in}(y) dy = \frac{2}{3} u_m, \tag{5}$$

where $u_m, D,$ and $h$ are the mean velocity, the cylinder diameter, and the height of the inlet boundary, respectively.

The governing equations (Eq.(3) and Eq. (4)) are solved in a segregated manner using incremental pressure correction scheme IPCS [42] based on FEniCS open source framework [43]. The FEniCS environment [43] utilizes finite element schemes with discontinuous Galerkin methods for spatial discretization [44]; this method has been adapted in pervious DRL-based flow control studies [13], [22], [23] to reduce the computation capacity during the training process of the deep neural networks DNNs. The IPCS method considers the continuity and the momentum equations separately in three steps [42]. The first step is solving the momentum equation (Eq. (4)) for a temporary value of the velocity ($u^*$) based on the pressure ($p^n$) and the velocity ($u^n$) that are known from the previous time-step using a midpoint finite difference scheme in time, as follows:

$$\frac{\partial u}{\partial t} = \frac{u^* - u^n}{\Delta t}, \tag{6}$$

The second step is calculating the pressure value $(p^{n+1})$ at the new time-step by solving a Poisson equation:

$$\nabla p^{n+1} = \nabla p^n - \frac{1}{\Delta t} \nabla . u^*, \quad (7)$$

then calculating the new time-step velocity $(u^{n+1})$ as follows:

$$u^{n+1} = u^* - \Delta t(\nabla p^{n+1} - \nabla p^n), \quad (8)$$

The drag and lift forces are integrated over the surface of the cylinder using the following formulas:

$$F_D = \int (\sigma . n_c) . e_x \, ds, \quad (9)$$

and

$$F_L = \int (\sigma . n_c) . e_y \, ds, \quad (10)$$

in which $e_x = (1,0)$ and $e_y = (0,1)$, $\sigma$ is the Cauchy stress tensor and $n_c$ is the unit vector normal to the surface of the cylinder. The drag and lift coefficients are derived from the drag and lift forces, respectively as follows:

$$C_D = \frac{2F_D}{\rho U^2 D}, \quad (11)$$

and

$$C_L = \frac{2F_L}{\rho U^2 D}, \quad (12)$$

The computational domain is discretized by an unstructured mesh generated using Gmsh [45]. The mesh consists of 4935 triangular cells and it is refined near the cylinder within the non-dimensional length of $X/D = 0.1 - 0.7$, as shown in Figure 6. To clarify mesh independence, two additional simulations are carried-out with grid consists of twice the amount of elements. As presented in Table 1, the discrepancy in drag coefficients is small, and the maximum value is within less than 1% from the reported values in [23], [27]. The maximum value of the lift coefficient is slightly larger than what is reported in [23], [27], but the discrepancy between the three values is small, and therefore, the mesh that consists of 4935 triangle cells is used thereafter considering that the drag reduction is the main focus of the current study. The time-step $(\Delta t)$ is set to 0.0125 s, corresponding to a non-dimensional time $U\Delta t/D = 2*10^{-3}$, to satisfy Courant number $U\Delta t/\Delta x < 1$. Similarly, three simulations with different time steps $\Delta t = 0.025, 0.0125, 0.00625$ are performed to clarify the time-step independency and the results show discrepancies with less than 1% corresponding to $C_{D_{max}} = 3.230, 3.223,$ and $3.218$, respectively.

(a)

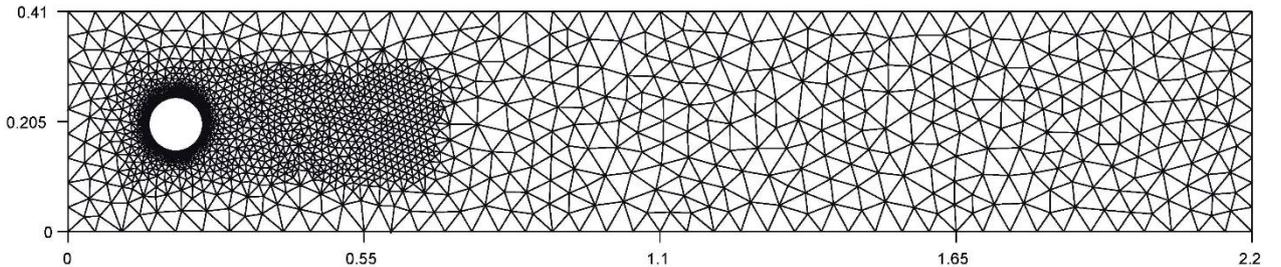

(b)

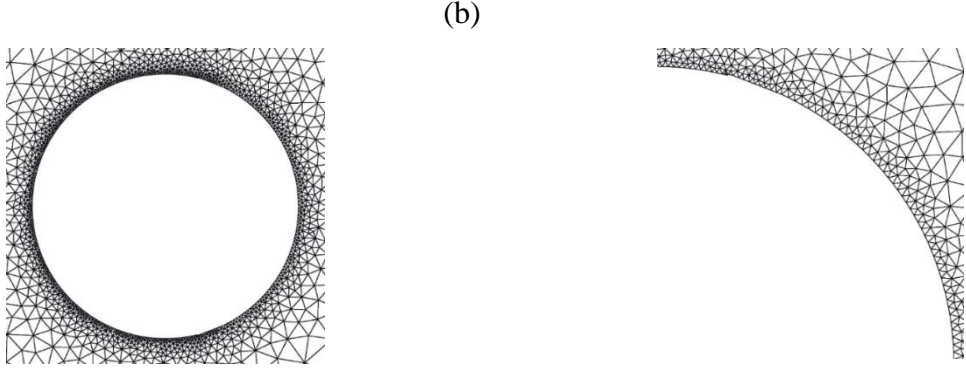

Figure 6: Description for the computational grid of the full domain (a) and near the circular cylinder (b), indication the mesh refinement near the wall of the cylinder.

Table 1 : Drag and lift coefficients for coarse, intermediate, and fine meshes compared to their counterparts from previous studies [23] and [27].

| Case | Mesh resolution | | $C_{D_{max}}$ | $C_{L_{max}}$ |
|---|---|---|---|---|
| Present | Coarse | 4935 | 3.223 | 1.058 |
| | Intermediate | 10423 | 3.242 | 1.061 |
| | fine | 20301 | 3.240 | 1.053 |
| H. Tang et al. [23] | | | 3.229~ 3.241 | 1.032~ 1.0758 |
| M. Schäfer et al. [27] | | | 3.220~ 3.240 | 0.990~ 1.010 |

## 3. Deep reinforcement learning (DRL) algorithm:

### 3.1. Short description for the deep reinforcement learning (DRL) paradigm:

A typical deep reinforcement learning (DRL) algorithm can be expressed in terms of three main paths: the action, the state, and the reward, as shown in Figure 7. The DRL agent interacts with an environment (here, FEniCS simulation framework) over a discrete number of time steps, and at each time step, the DRL agent receives information about the current state $(s_t)$ then selects an action $(a_t)$ from the action-space corresponding to the policy $(\pi)$. After each action, the agent receives information about the new state $(s_{t+1})$ and the reward $(r_t)$ until the agent approaches a terminal. The goal is to find an optimal policy $(\pi_{opt})$ that maximizes the expected return from each state $(R_t)$ which is defined as follows:

$$R_t = \sum_{k=0}^{\infty} \gamma^k r_{t+k}, \qquad (13)$$

in which $r$ is the reward, and $\gamma \in [0,1]$ is the discount factor. After the DRL agent executes an action, the reward is obtained as the time-averaged drag coefficient penalized by the time-averaged lift coefficient as follows:

$$r_t = |C_D|_T - \alpha |C_L|_T, \qquad (14)$$

in which $T$ is the period of one Kármán vortex cycle (here, $T_K = 20.49$ s), and the parameter $\alpha$ is set to 0.2 similar to that in [13]. The lift penalization in Eq.(14) is used to prevent the neural network from cheating and to provide a practical estimation of the reward; more details about the lift penalization can be found in [13].

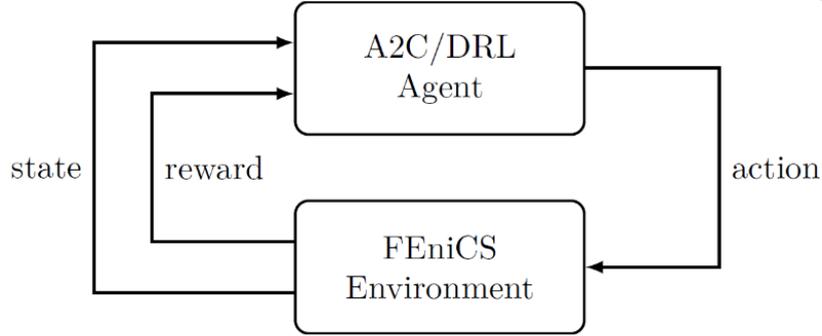

Figure 7: Schematic diagram of a typical reinforcement learning algorithm.

The ability of observing the state of the environment remains a challenging task in reinforcement learning research [46], and generally, the state is detected by a partial observation of the environment. In the literature, three methods are suggested for assessing the flow-state in active flow control (AFC) problems: (1) concatenation between drag and lift coefficients [47], (2) a set of velocity probes positioned downstream the bluff-body [13], and (3) a set of pressure probes located on the surface of the bluff-body [48]. Similar to the latter technique in [48], ten pressure probes located on the surface of the cylinder are used in the current study to send the feedback observations to the deep neural networks (DNNs), as depicted in Figure 3. The selection of the location and the appropriate number of the feedback probes is a widely explored area that has been discussed in pervious AFC studies [13], [49], but this is beyond the scope of the current study and it is left for future work.

### 3.2. Architecture of the advantage actor-critic (A2C) algorithm:

The advantage actor-critic (A2C) algorithm [26] belongs to the policy gradient family [50], and it consists of two deep neural networks: the policy network (known as the actor) and the value-function network (known as the critic). As shown in Figure 8, the actor network is used to select actions from the action-space, while the critic network is used to criticize the actions that are selected by the actor. The training process is always on-policy, and the critique takes the form of a TD error (here, it is TD(0)).

As shown in Algorithm 1, the state-value function ($V^\pi$) is used as a baseline function (i.e. an advantage function) for calculating the TD error ($\delta$) to reduce the variance estimate of the policy gradient, as discussed in [26]. The A2C algorithm is implemented in Python based on the deep learning package Tensorflow [51] backend to Keras. All networks are feed-forward neural networks consist of two dense layers, with a width of 128 and 64, respectively. A discount factor $\gamma = 0.99$ and Adam optimizer with a learning rate of 0.001 are used. All networks parameters are modified by hand-tuning from a baseline algorithm (https://github.com/philtabor/Actor-Critic-Methods-Paper-To-Code), which is created to control the well-known CartPole problem.

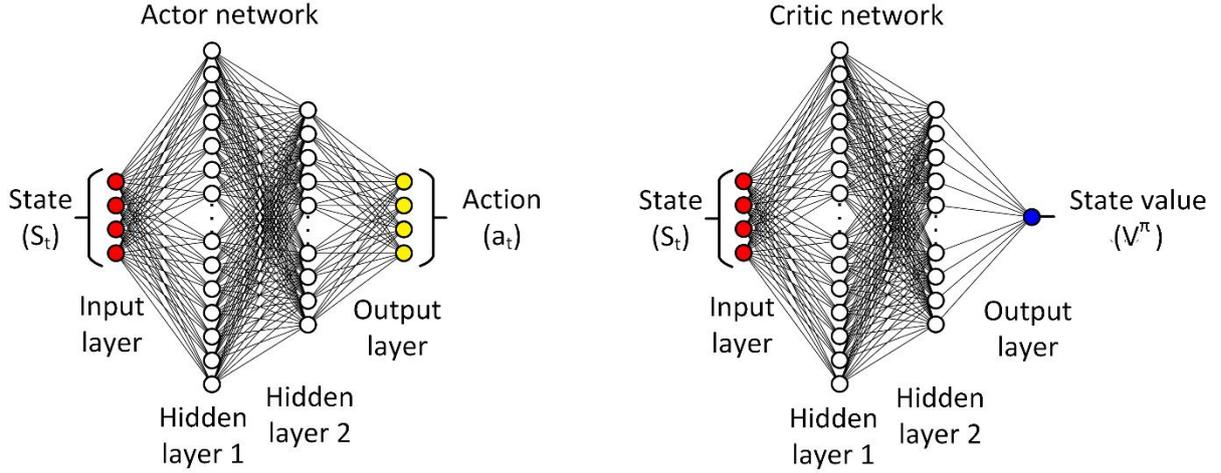

Figure 8: Description of the critic network (right) and the actor network (left). The two networks consist of four layers: the input layer, the first and the second hidden layers, and the output layer. The input and the two hidden layers are identical in both networks, but the output layer is different. In the actor network, the output layer consists of 4 neurons corresponding to the number of actions, while in the critic network the output layer includes only one neuron for the state value ($V^\pi$).

Algorithm 1: Advantage actor-critic (A2C)
Initialize randomly the policy network parameters $\theta$
Initialize randomly the value-function network parameters $\omega$
Loop for each episode:
    Initialize $s_0$ and set $t = 0$
    Loop while $s$ is not terminal (for each time step $t$):
        Sample $a_t \sim \pi_\theta(a|s_t)$
        Execute $a_t$, observe $s_{t+1}, r_t$
        $\delta = r_t + \gamma\, V_\omega(s_{t+1}) - V_\omega(s_t)$
        update value function parameter: $\omega = \omega + \alpha_\omega \gamma^t \delta \nabla V_\omega(s_t)$
        update policy parameter: $\theta = \theta + \alpha_\theta \gamma^t \delta \nabla \log \pi_\theta(a_t|s_t)$
        $t = t+1$

## 4. Results and discussion:

An efficient control strategy is obtained from manipulating the non-dimensional burst frequency of the two plasma actuators by means of the advantage actor-critic (A2C) algorithm. Four non-dimensional burst $f^+ = 0.5, 1.0, 1.5$, and $2.0$, corresponding to plasma burst frequencies $f_b = 0.08$ Hz, 0.16 Hz, 0.24 Hz, and 0.32 Hz are investigated at a constant body force per unit volume $F = 0.7$ mN/m$^3$ and a fixed duty-cycle $\tau = 85\%$. By applying the methodology discussed in the previous section, the DRL agent is able to learn a control strategy that achieves a drag reduction of 22.6 % after an average value of 96 training episodes. The effectiveness of the control strategy is evaluated by investigating three parameters, namely the drag coefficient, the lift coefficient, and the average velocity magnitude at the wake region. The fast Fourier transformer (FFT) is also utilized to validate the results and illustrate the effect of the plasma actuators on the flow field.

The advantage actor-critic (A2C) algorithm that is utilized for training the deep neural networks DNNs is episode-based which means that the training process is divided into a number of training sequences [50]. Firstly, the numerical simulations are carried-out under the uncontrolled conditions (without actuation) for a non-dimensional time $T^* = 24.83$ until a fully-developed unsteady wake (i.e. Kármán vortex street) is detected. Then, the initial state is used to start the following steps in each episode until the target is reached. At the beginning of the training process, the weights of all neural networks are initialized randomly and the actions are selected from the action-space according to the probabilities of each action in the actor network, then the same process is repeated until the networks converge. Each action takes time that equals to one Kármán shedding cycle (here $T_K^* = 3.28$), and the drag and lift coefficients are estimated by averaging over the same period as discussed in the previous section.

Figure 9 shows the progress of the training process in which three training sessions, with different random seeds, are performed to clarify the robustness of the control strategy. At each training session, the DRL agent follows an independent path, but eventually all paths reach the same final policy. The agent's target is achieving the maximum accumulated reward within the minimum number of actions, and the number of actions varies according to the number of the visited states at each episode. In the current study, a limited number of ten states per each episode is set to reduce the computation time and to protect the DRL agent from getting involved into an infinite number of actions. The training process takes place in an average value of approximately 96 episodes, but the process has been pursued up to 250 episodes to confirm the stability of the learning strategy.

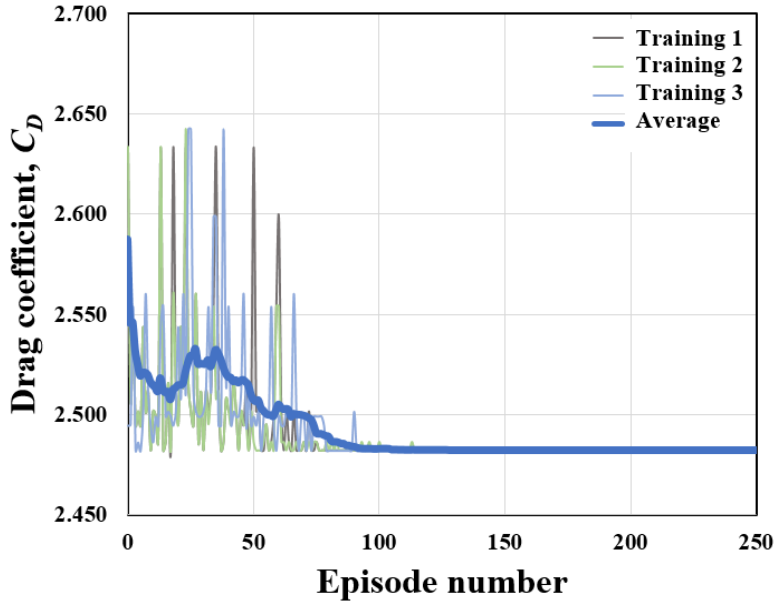

Figure 9: Description of the training process showing the instantaneous and the average drag coefficients of three training sessions. At each training session, the deep neural networks DNNs are initialized using different random seeds, and the final control strategy is achieved after approximately 76, 114, and 91 for the first, the second, and the third sessions, respectively. The instantaneous drag coefficient is considered as the overall drag coefficient that is obtained after the end of each episode, and the average drag coefficient is estimated by averaging the instantaneous drag coefficients of each ten episodes.

A comparison between the flow undergoing control and the uncontrolled flow in terms of the mean velocity and the mean pressure distributions is presented in Figure 10 and Figure 11, respectively. It can be observed from Figure 10 that the Kármán vortex street is significantly altered and the velocity fluctuations become less strong near the upper and the lower boundaries. Figure 10 also indicates that the size of the recirculation-zone downstream the cylinder in the controlled case become elongated and less tapered compared to the uncontrolled case. This correlation between the increase in the length of the recirculation-zone and the reduction in the drag coefficient is a well-known fact that has been discussed in previous flow control studies [49], [52]. On the other hand, Figure 11 indicates that there is a strong recovery on the pressure distribution downstream the cylinder consistent with the drag reduction. It can also be observed that the recirculation bubbles downstream the cylinder in the uncontrolled case become clearly visible until a non-dimensional length of $X/D = 1.65$, while in the controlled case the recirculation bubbles are less dense and they can be barely seen after a non-dimensional length of $X/D = 0.5$.

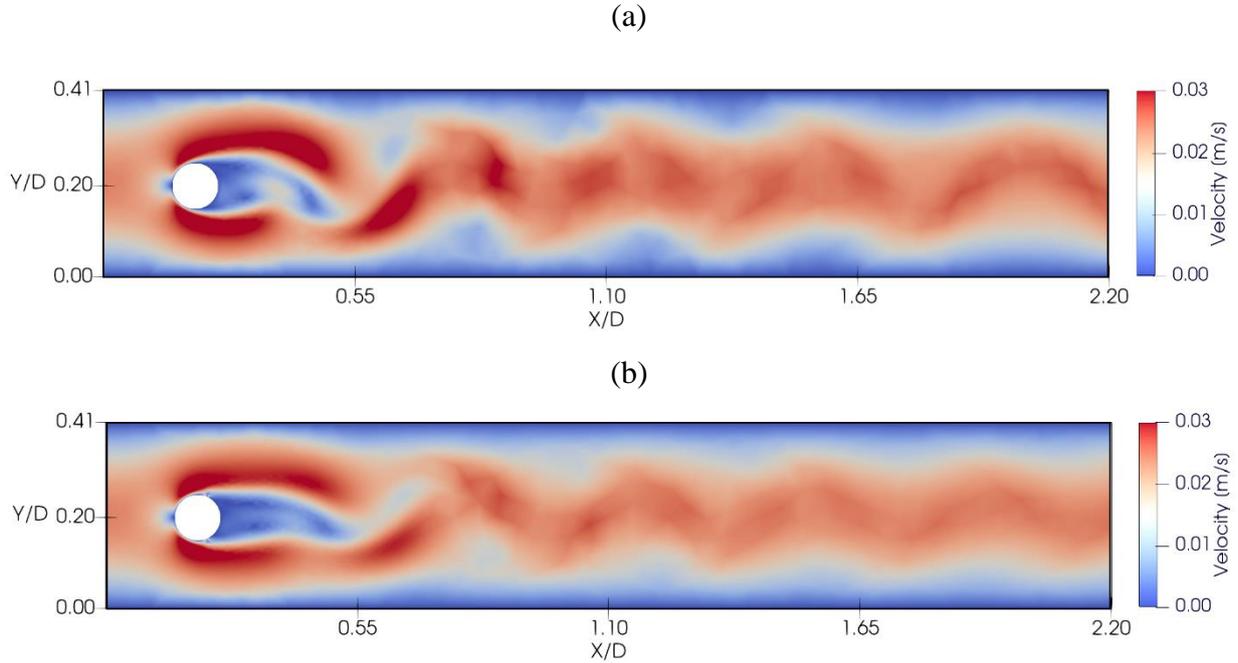

Figure 10: Snapshots of the flow field show a comparison between the uncontrolled case (a) and the controlled case (b) for the distribution of the mean velocity magnitude across the full computational domain. The uncontrolled case (a) and the controlled case (b) are presented after a time period of $T = 155.2\ s$ and $T = 942.45\ s$, respectively

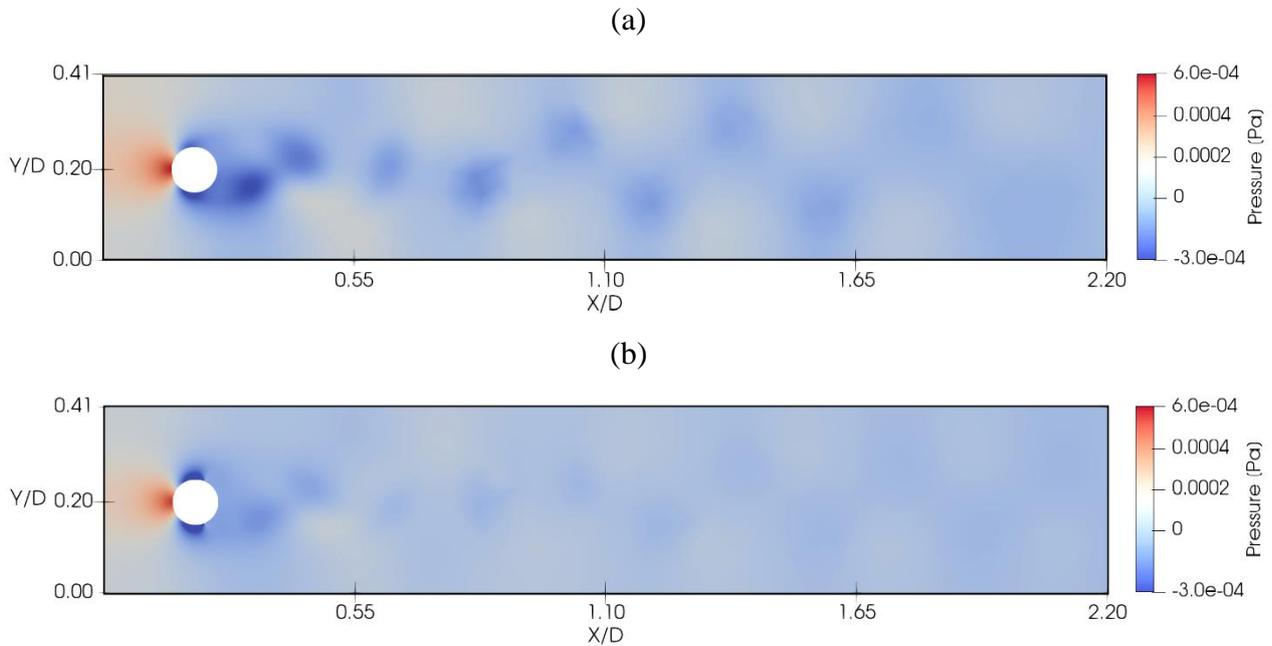

Figure 11: Snapshots of the flow field show a comparison between the uncontrolled case (a) and the controlled case (b) for the distribution of the mean pressure across the full computational domain. The uncontrolled case (a) and the controlled case (b) are presented after a time-period of $T = 155.2\ s$ and $T = 942.45\ s$, respectively.

To further investigate the effect of the control strategy on the flow-field, the time-series of the mean velocity magnitude are recorder from five locations downstream the cylinder at $X/D = 0.9$ and $Y/D = 0.15, 0.18, 0.20, 0.22,$ and $0.25$. Figure 12 compares the average value of the mean velocity magnitude, which is calculated from the five time-series, at the controlled case to its counterpart from the uncontrolled case. A significant reduction in the mean velocity magnitude and in the mean velocity fluctuations by respectively 13% and 30% is obtained. In addition, the natural vortex shedding frequency of the controlled flow and the uncontrolled flow is estimated by applying the fast Fourier transformer (FFT) to the average velocity signals after subtracting their mean values to make the results clearly visible. The Strouhal number ($St = fD/U$) of the uncontrolled case is estimated as $St = 0.305$ which is in a good agreement with less than 1% from what is reported in [23], [27]. Figure 12 also indicates that the Strouhal number ($St$) increases by approximately 10% after applying the control strategy, which is consistent with what is discussed in the previous DRL-AFC problem in [23] where synthetic jets are utilized to control the flow around a 2-D circular cylinder at a diameter based Reynolds number $Re_D = 100$.

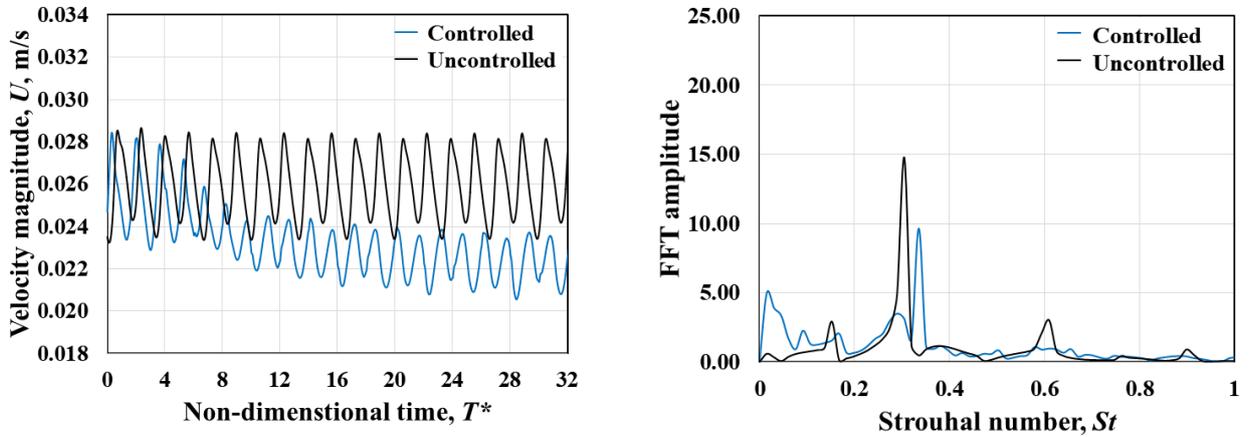

Figure 12: Comparison between the uncontrolled case and the controlled case showing the time history (left) and the fast Fourier transformer (right) of the average velocity magnitudes that are recorded from five velocity probes located at $X/D = 0.9$ and $Y/D = 0.15, 0.18, 0.20, 0.22,$ and $0.25$. The time signals are recorded for a time-period equals to approximately ten Kármán shedding cycles (here $T_K^* = 3.28$).

Figure 13 compares the temporal development of the drag and lift coefficients at the controlled case with the uncontrolled case. The figure shows a significant reduction on the drag coefficient after applying the plasma actuators as an average drag reduction of approximately 22.6% is reached, with a maximum value of $C_D = 2.01$ corresponding to a drag reduction of 37 %. In addition to the drag reduction, the plasma actuators are able to suppress the lift fluctuations to more than one-quarter from a maximum value of $C_l = 1.028$ to $C_l = 0.247$. As shown in Figure 14, the lift coefficient ($C_L$) changes periodically with a frequency that is equal to the natural vortex shedding frequency ($St$= 0.305), while the drag coefficient ($C_D$) changes with a frequency that is as twice as the natural vortex shedding frequency ($St = 0.616$). These observations are excepted and they are similar to what is discussed in [13], [53].

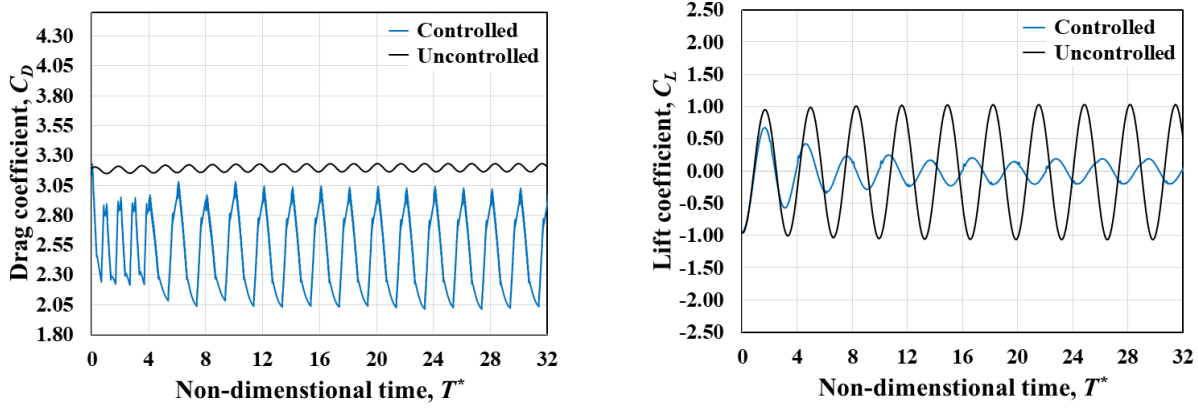

Figure 13: Comparison between the uncontrolled case and the controlled case in terms of the evolution of the instantaneous drag coefficient $C_D$ (left) and the instantaneous lift coefficient $C_L$ (right) for a time-period equals to approximately ten Kármán shedding cycles (here $T_K^* = 3.28$).

Furthermore, the fast Fourier transformer (FFT) of the drag coefficient in the controlled case indicates that the DRL agent utilized only two non-dimensional burst frequencies, namely $F^+ = 0.5$ and $1.0$, for achieving the flow control strategy. Interestingly, this control strategy that is learned by the DRL agent results in a drag reduction that is higher than what is obtained from a regular actuation using a constant non-dimensional burst frequency. More precisely, performing a modulation using a constant $F^+= 0.5$ and $F^+= 1.0$ yields to a drag reduction of respectively 21.8 % and 21.1%, while the DRL agent is able to achieve a drag reduction of 22.6% resulting from utilizing both $F^+= 0.5$ and $F^+= 1.0$.

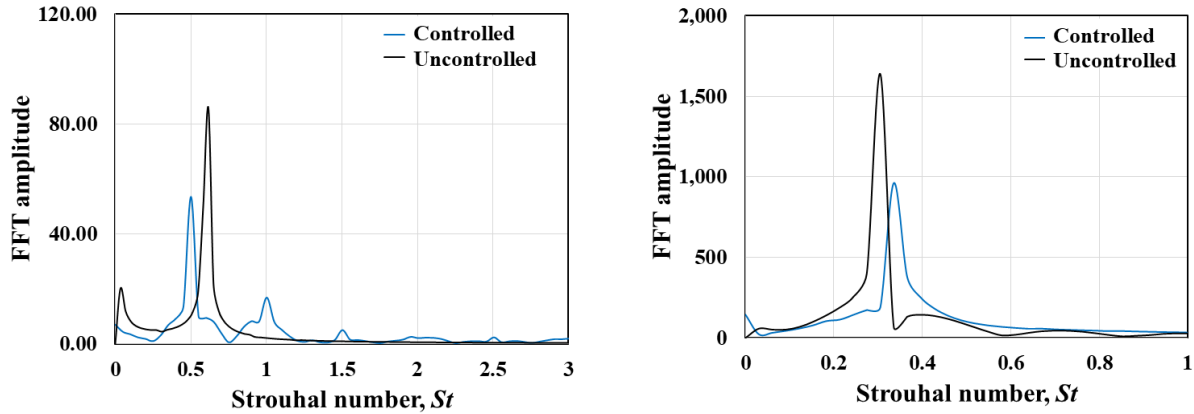

Figure 14: Comparison between the Fast Fourier transformer (FFT) of the drag coefficient (left) and the lift coefficient (right) for the controlled case and the uncontrolled case estimated for a time-period equals to ten Kármán shedding cycles (here $T_K^* = 3.28$). To make the comparisons clearly visible, the FFT amplitude of the drag coefficient for the controlled case and the FFT amplitude of the lift coefficient at the uncontrolled case are divided by 15 and 2, respectively.

Figure 15 shows a close look at the effect of the plasma actuators on the flow-field, indicating the initialization of two small-scale counter-rotating vortices near the upper and the lower end of the cylinder. Each time the plasma actuators are activated, these vortex pairs are released from the end of the cylinder and mix chaotically with the alternating vortex shedding. The counter-rotating motion of these vortex pairs near the separation point brings high-momentum fluid to the near wall region helping the flow to withstand the adverse pressure gradient, and thus, delay the flow separation [3], [31]. This effect of the plasma actuators on the flow reattachment can be clearly observed in Figure 16, which shows that the flow-separation angle is shifted by at least 10° after using the plasma actuator. Figure 16 also shows that the size of the recirculation bubbles decreased significantly in the actively controlled case, which illustrates the increase of the natural vortex shedding frequency that has been discussed previously in Figure 12.

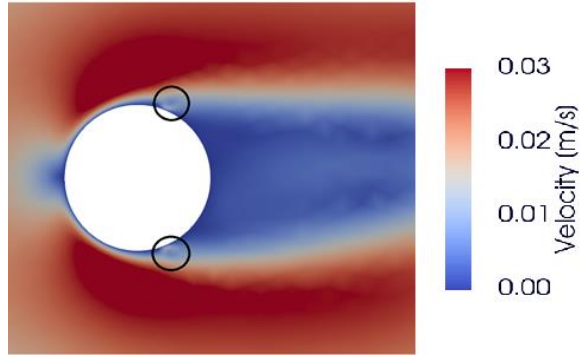

Figure 15: A snapshot of the flow field showing the mean velocity magnitude near the cylinder indicating the initialization of the two counter-rotating vortices that are generated by the plasma actuators.

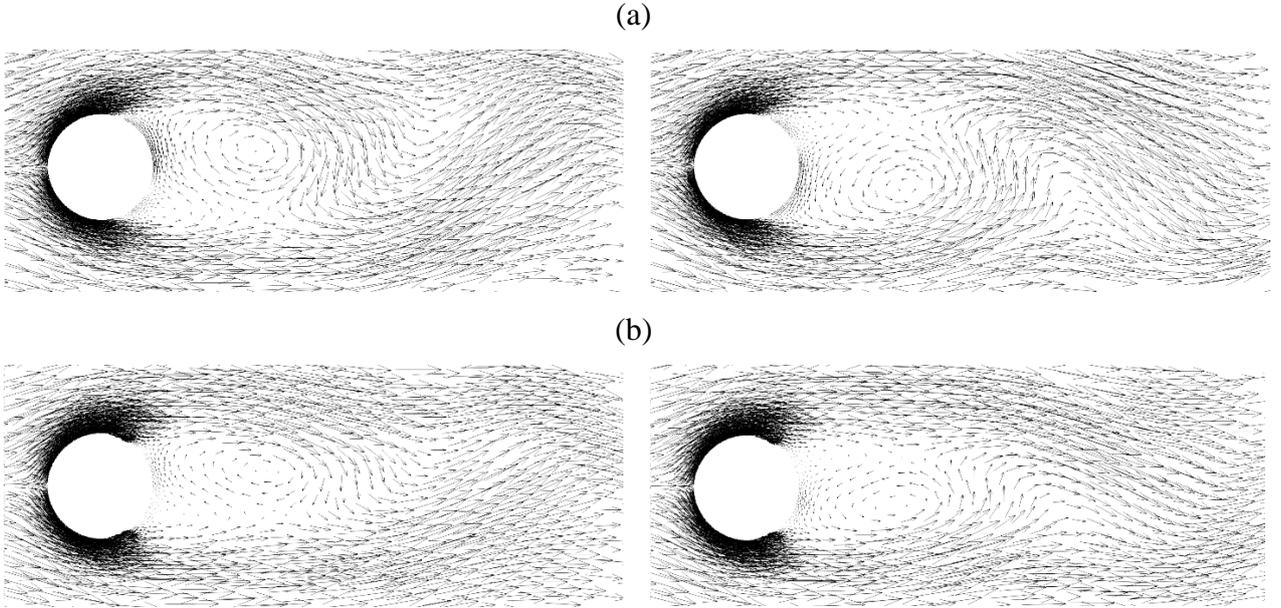

Figure 16: Instantaneous velocity vectors for the uncontrolled case (a) and the controlled case (b), indicating the effect of the plasma actuator on the flow reattachment as well as on the size of the recirculation bubble. In the uncontrolled case, the upper and the lower vortex shedding are shown after a period of respectively $T = 106.15\ s$ and $T = 117.15\ s$, and in the controlled case the upper and lower vortex shedding are presented after a period of $T = 842.108\ s$ and $T = 852.108\ s$, respectively.

## 5. Conclusion:

The current study demonstrates the robustness and feasibility of the deep reinforcement learning (DRL) paradigm, and more specifically the advantage actor-critic (A2C) algorithm, in performing active flow control (AFC). Flow control strategies are investigated for flow past a 2-D circular cylinder at a diameter-based Reynolds number $Re_D = 100$ via manipulating the non-dimensional burst frequency of two symmetric plasma actuators located at angle $\theta = \pm 115°$ on the surface of the cylinder . Numerical simulations are carried-out to solve the 2-D unsteady incompressible Navier-Stokes equations using the incremental pressure correction scheme (IPCS) based on FEniCS open source framework. Four non-dimensional burst frequencies $f^+$ = 0.5, 1.0, 1.5, and 2.0, corresponding to plasma burst frequencies $f_b$ = 0.08 Hz, 0.16 Hz, 0.24 Hz, and 0.32 Hz, are tested at a fixed actuation body force ($F$) and constant duty-cycle ($\tau$). During the training process, the deep neural networks (DNNs) receive feedback observations from ten pressure sensors located on the surface of the cylinder, and the exploration of the control strategy is achieved through optimizing a reward function that is defined from the fluctuations of the drag and lift coefficients.

The advantage actor-critic (A2C) agent is able to learn a control strategy that achieves a drag reduction of 22.6% after an average value of 96 episodes. Three training sessions with different random seeds are performed to clarify the robustness of the control strategy and it has been shown that all sessions reached the same final policy. A close look at the flow field shows that the plasma actuators generate two small-scale counter-rotating vortices near the upper and the lower end of the cylinder. These vortex pairs mix chaotically with the alternating vortex shedding downstream the cylinder helping the flow to withstand the adverse pressure gradient, and thus, delay the flow separation. Furthermore, it has been shown that the size of the recirculation zone downstream the cylinder becomes elongated and narrower when the control strategy is applied consistent with the drag reduction.

In addition, a significant decrease in the wake velocity magnitude and fluctuations is observed, and the fast Fourier transformer (FFT) shows that the natural vortex shedding frequency of the actively controlled case is increased by approximately 13% compared to the uncontrolled case. Furthermore, the fast Fourier transformer (FFT) of the drag coefficient indicates that the DRL agent selected only two non-dimensional burst frequencies, namely $f^+$ = 0.5 and 1.0, for achieving the control strategy. Interestingly, the DRL agent's control strategy yields to a drag reduction that is higher than that from a regular actuation using constant non-dimensional burst frequencies, which indicates the great capabilities of deep reinforcement learning algorithms on discovering novel control strategies that are beyond the human capabilities.

In order to facilitate the further development of DRL/A2C algorithm, All codes that have been used in the current study are released as open-source on the GitHub (https://github.com/maelhawary/DRL_for_flow_control)

## Conflict of interest:

The author declares that there is no conflict of interest.